\documentclass{article}
\usepackage{graphics,graphicx}

\title{\bf Motion of a particle and the vacuum\footnote{
{\it Physics Essays}, {\bf 6}, no. 4, pp. 554-563 (1993)}}

\author{{\bf Volodymyr Krasnoholovets and Dmytro Ivanovsky}   \\
 {} \\
Institute of Physics, National Academy of Sciences,   \\ Prospect
Nauky 46,  UA-03650 Ky\"{\i}v 39, Ukraine}

\date{Received 16 April 1992}

\begin{document}
\maketitle

\begin{abstract}

\hspace*{\parindent}

We propose the deterministic dynamics of a free particle in a
physical vacuum, which is considered as a discrete (quantum)
medium. The motion of the particle is studied taking into account
its interactions with the medium. It is assumed that this
interaction results in the appearance of special virtual
excitations, called "inertons," in the vacuum medium in the
surroundings of the canonical particle. The solution of the
equation of motion shows that a cloud of inertons oscillates
around the particle with amplitude $\Lambda=\lambda v_0/c$, where
$\lambda$ is the de Broglie wavelength, $v_0$ is the initial
velocity of the particle, and $c$ is the initial velocity of the
inertons (velocity of light). This oscillating nature of motion is
also applied to the particle, and the de Broglie wavelength
$\lambda$ becomes the amplitude of spacial oscillations. The
oscillation frequency $\nu$ is given by the relation $E=h\nu$. The
connection of the present model with orthodox nonrelativistic wave
mechanics is analyzed.

\vspace{2mm}


\end{abstract}

\vspace{5mm}
\section{Introduction}

\hspace*{\parindent}

The concepts of condensed media physics penetrate more and more
deeply into the models of elementary particles [1,2] and into the
models of the physical vacuum [1, 2-5]. Specifically, in Ref. 3
the vacuum is considered as  an elastic medium; more
fundamentally, the possibility of simulating the vacuum with the
form of a quantum crystal is demonstrated in Ref. 5. A subquantum
medium is also presented in papers studying the problem of the
causal interpretation of quantum mechanics [6].

The present paper proposes a deterministic dynamics of a free
elementary particle in a physical vacuum considered as a discrete
(quantum) substance. The "molecules" of this substance, which may
be called superparticles, form a peculiar elastic medium. The
concept of the mass of the particle is introduced, and its motion
is studied, taking into account its interaction with the vacuum
medium. It is suggested that this interaction results in the
appearance of special virtual excitations of the medium. The
relationship between the present nonrelativistic model and
orthodox nonrelativistic quantum mechanics is analyzed.

\vspace{4mm}

\section{A particle in a vacuum medium}

\vspace {2mm} \hspace*{\parindent}

A particle will be taken to mean one of multiple states of a
superparticle corresponding to a lepton, for example, an electron.
It is reasonable to assume that the size of a superparticle in the
degenerate state is close to the value $10^{-28}$ cm (the size
required by the grand unification of interactions). At such scales
gravitational effects can become extremely substantial [4,5]. One
of the manifestations of microgravitation can be can be an effect
similar to the state of a polaron in a crystal, that is,
deformation of the vacuum in the neighborhood of the particle.
Clearly, the simplest of the reasons that can cause such
deformation is a deviation of the typical dimension of the
particle creation from the dimension of the initial degenerate
superparticle. Figure 1 pictorially illustrates this.
\begin{figure}
\begin{center}
\includegraphics[scale=0.5]{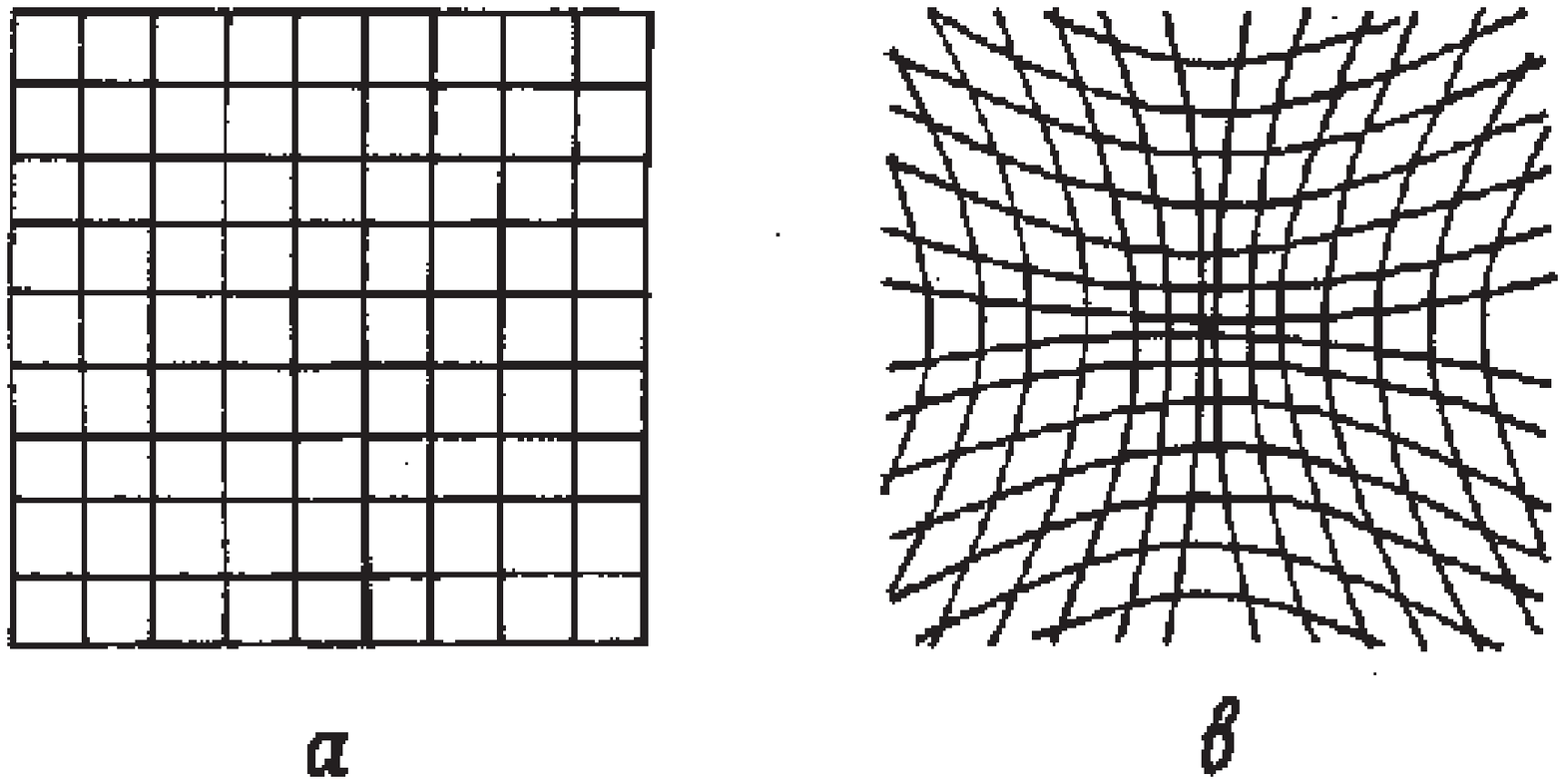}
\caption{ Cells of degenerate vacuum medium ({\it a}) (left) and
their deformation in the neighborhood of the particle ({\it b}).}
\label{Figure 1}
\end{center}
\end{figure}
A vacuum medium is shown in Fig. 1, with  each cell occupied by a
superparticle in the degenerate state and interacting only with
the nearest neighbors. If the superparticle possesses elastic
properties, then the vacuum medium naturally should deform in the
neighborhood of the particle at the creation of the particle
(i.e., when the size of one of the superparticles varies from the
initial value $R_0$ to the fixed value $R_{\rm part}$, for
example, $R_{\rm part}< R_0$) [Fig. 1({\it b})].

In the initial state [Fig. 1({\it a})], the special position of
the medium cells is characterized by Cartesian coordinates $X^i$,
but with the creation of a particle, the new symmetry [Fig. 1({\it
b})] is already described by generalized coordinates $Q^k$ which
are functions of $X^i$. This distortion of boundary lines of the
vacuum medium cells is similar to gravitational field lines.
Therefore, it can be said that the particle is in a gravitational
potential which defines the distortion of space.

In the general theory of relativity, the components of the metric
tensor $g_{ij}$ act as the gravitational potential, but in our
model, quantities $g_{ij}$ characterize the variation of the
equilibrium position of superparticles. Identifying the
deformation of the vacuum medium caused by the particle with
gravitational potential of the particle, we can thus relate the
stationary relative deviation of the dimension of the
superparticle from the initial value $R_0$ to the appearance of
mass in them. Taking into account the three dimensionality of
space, the mass of the superparticles should be related to the
relative variation of volume: $m(Q^k)= {\rm const} R_0^3/R^3_{\rm
part}$ (but in the general case the ratios between volumes, i.e.,
between mass $m$ and $M$, are tensor quantities).

It should be noted that the vacuum may be regarded as a special
crystal only within the deformation "coat" (or distance) formed in
the discrete vacuum medium due to particle creation
(superparticles are characterized by mass only within this
region). This crystal should obviously possess some features
typical of a solid crystal. The vacuum medium is found in a
degenerate state beyond the limits of the "coat."

\vspace{4mm}

\section{The hypothesis of motion}

\vspace{2mm} \hspace*{\parindent} The motion of a large polaron in
a solid (see, e.g., Ref. 7) comprises the motion of the charge
carrier and that of the polaron "coat." Similarly, in a vacuum
medium a particle in a motion will also pull behind it the vacuum
deformation "coat" (Fig. 1) will remain invariable at any point of
its location (in order that this takes place, the rate of
adjustment of superparticles naturally should be higher than the
velocity of the particle). The particle moving in a superdensely
packed medium naturally cannot move freely; it certainly should
experience friction losing the velocity as this takes place. As a
result of the interaction with the medium, the particle emits
(absorbs) the elementary virtual excitations of this medium. It is
reasonable to relate the initial speed of these excitations with a
value on the order of or equal to the velocity of light $c$, but
the velocity of the particle $v_0< c$. This inequality and the
availability of the deformation potential created by a particle in
a vacuum medium allow one to assume in principle the possibility
of nonstationary motion whose essence can be reduced to the
following.

Emitted (at an angle to the trajectory) excitations are ahead of
the particle, but they are gradually retarded by its potential;
then, having reached the boundary deformation coat-degenerated
vacuum, return again to the particle (i.e., to the center of the
potential maintaining them) and transmit to it the lost speed.
However, to the moment of the return of an excitation to the
particle, the latter has already shifted forward from the point
where the excitation was emitted. Therefore, the excitations
emitted by the particle ahead themselves will return to it from
behind so that the father particle will be moved forward in a
direction that is defined by the initial velocity vector ${\vec
v}_0$ (Fig. 2). Let us study this motion mathematically under the
assumption that external fields (gravitational, etc.) are absent.

\begin{figure}
\begin{center}
\includegraphics[scale=0.5]{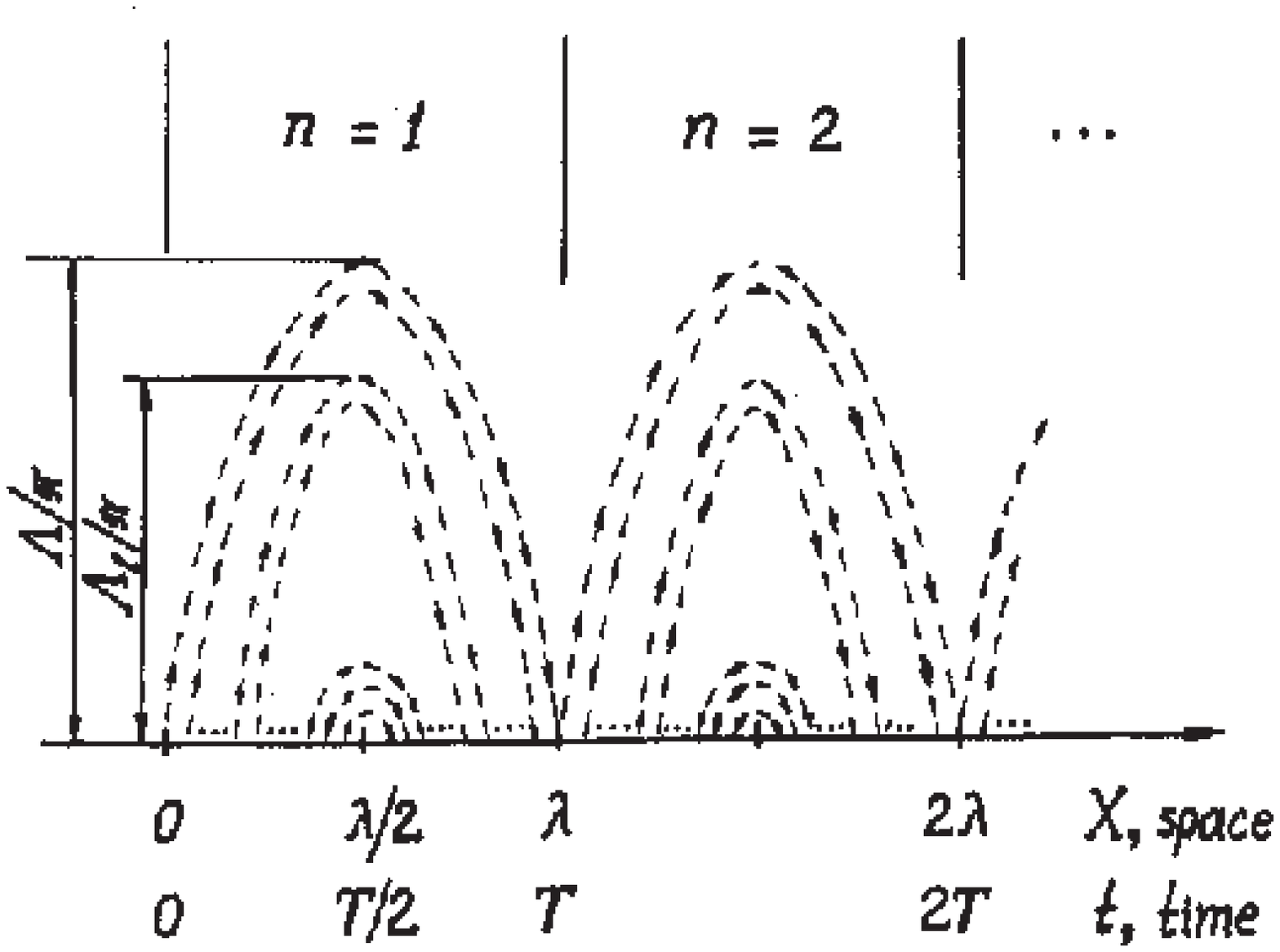}
\caption{Trajectories of motion of a particle and excitations.}
\label{Figure 2}
\end{center}
\end{figure}

\section{Lagrangian and the equation of motion}

\vspace{2mm} \hspace*{\parindent} Let us proceed from the
Lagrangian
\begin{eqnarray}
L&=&\frac 12 g_{ij}{\dot X}^i(t){\dot X}^j(t) + \frac 12
\sum^{N-1}_{l=0} {\tilde g}^{(l)}_{ij} {\dot x}^i_{(l)}(t_{(l)})
{\dot x}^j_{(l)}(t_{(l)})    \nonumber   \\
&-&\sum_{l=0}^{N-1}\frac {\pi}{T} \delta_{t-\Delta t_{(l)},
t_{(l)}} \Bigl\{ X^i(t)
 \sqrt{g_{ir}({\hat A}^{-1}{\tilde
g}^{(l)}_{rj})_0} \ \  {\dot x}^j_{(l)}(t_{(l)})  \\  \nonumber
&+&{\dot X}^{i}(t)\Big|_{t=0} \sqrt{g_{ir} ({\hat A}^{-1}{\tilde
g}^{(l)}_{rj})_0} \ \  x^j_{(l)}(t_{(l)})  \Bigr\} \label{1}
\end{eqnarray}
The first term characterizes the particle, the second term
characterizes the ensemble of $N$ excitations, emitted by the
particle, and the third characterizes the contact interaction
between the particle and the ensemble. $X^i$ and ${\dot X}^i$ are
the $i$th components of the position and the velocity of the
particle, respectively; $g_{ij}$ is a metric tensor generated by
the particle in three-dimensional space. Indices $l$ correspond to
the number of the respective excitation; they are enclosed in
parentheses to distinguish them from indices [$i, j$, and $r$ in
Eq. (1)] describing components of vector and tensor quantities;
$x^j_{(l)}$ and ${\dot x}^i_{(l)}$  are components of position and
velocity of the $l$th excitation; ${\tilde g}^{(l)}_{ij}$ is the
metric tensor generated by the $l$th excitation in
three-dimensional space (it describes local deformation in the
neighborhood of the excitation); $1/T_{l}$ is the frequency of
collisions of the particle with the $l$th excitation; Kronecker's
symbol $\delta_{t-\Delta t_{(l)}, t_{(l)}}$ provides the agreement
of proper times of the particle $t$ and the $l$th excitation at
the instant of their collision ($\Delta t_{(l)}$ is the time
interval after the expiry of which, measuring from the initial
moment $t=0$, the moving particle emits the $l$th excitation).
Operator ${\hat A}^{-1}$ appearing in the interaction energy
characterizes the rotation of three-dimensional space around axis
$X^i$; matrix $A$ belongs to the group of rotation of
three-dimensional space around axis $X^i$; matrix $A$ belongs to
the group of rotation SO(3). This rotation of space eventually
results in the motion of excitation metric ${\tilde
g}^{(l)}_{rj}\rightarrow {\tilde g}^{(l)}_{r+\alpha, j}$ (with
regard to the cyclic permutation index $\alpha$ in the same manner
as indices $r$ and $j$ take on values 1, 2, 3), that is, operator
${\hat A}^{-1}$ acts according to the following rule:
\begin{eqnarray}
g_{ir}\bigl( {\hat A}^{-1} {\tilde g}^{(l)}_{rj} \bigr)_0 &=&
g_{ir}\Bigr( \frac{1}{A}\frac{\partial z^k_{(l)}}{\partial
x^r_{(l)}}\frac{\partial z^k_{(l)}}{\partial x^j_{(l)}} \Bigr)
\Big|_{x_{(l)}=0}   \nonumber   \\ &=&g_{ir}\Bigr( \frac{\partial
z^k_{(l)}}{\partial A x^r_{(l)}}\frac{\partial z^k_{(l)}}{\partial
x^j_{(l)}}\Bigr) \Big|_{x_{(l)}=0}= g_{ir}\bigl( {\tilde
g}^{(l)}_{r+\alpha, j}\bigr)_0
  \label{2}
\end{eqnarray}
the radius vector $x_{(l)}(x^j_{(l)})$ of the $l$th excitation is
measured from the radius vector of the particle $X(X^i)$. For
example, if rotation takes place in Euclidean space around axis
$X^1$, then (see Ref. 8)
\begin{eqnarray}
A=\left( \matrix{1  &    0       &      0             \cr
            0    & \cos\varphi   &      \sin\varphi   \cr
            0    & -\sin\varphi  &   \cos\varphi      \cr } \right)
 \label{3}
\end{eqnarray}
where  $\varphi$ is the angle of rotation. Thus operator ${\hat
A}^{-1}$ acting according to rule (2) provides the change from the
space of the particle motion to the space of excitation motion,
that is, it shifts the excitations to trajectories different from
the trajectory of the particle (see Fig. 2). In the subsequent
discussion we shall use the notation
\begin{equation}\label{4}
  {\hat B}^{l}_{ij}=\sqrt{g_{ir} \bigl( {\hat A^{-1}}{\tilde g}^{(l)}_{rj}
   \bigr)\big|_0} \ .
\end{equation}
Let us write the Euler-Lagrange equations
\begin{equation}\label{5}
 d (\partial L / \partial {\dot Q}^k)/d t - \partial L /\partial Q^k = 0,
\end{equation}
where for the particle $$ Q^k = X^k (t_{(l)}+ \Delta
t_{(l)})\equiv X^k_{(l)}(t_{(l)}+
  \Delta t_{(l)}) \eqno(5a)
$$
 and for excitation
$$
  Q^k = x^k_{(l)}(t_{(l)}) \eqno(5b)
$$ we take into account the availability of the symbol
$\delta_{t-\Delta t_{(l)}, t_{(l)}}$ in the potential energy in
Eq. (1)]. From Eq. (5), taking into account Eqs. (1) and (3), we
obtain the equation of extremals:
 $$
{\ddot X}^s_{(l)} + \Gamma_{ij}^s {\dot X}^i_{(l)}{\dot X}^j_{(l)}
+ \frac{\pi}{T_{(l)}}g^{ks}\Bigl( \frac{\partial {\hat
B}^{(l)}_{ij}}{\partial X^k_{(l)}} \Bigr) \bigl( X^i_{(l)}{\dot
x}^j_{(l)} + {\dot X}^i_{(l)}\Big|_{t_{(l)}=0} x^j_{(l)} \bigr)
 $$
  $$ +\frac{\pi}{T_{(l)}} g^{ks}{\hat B}^{(l)}_{kj}{\dot x}^j_{(l)}
= 0; \eqno(6)
  $$
  $$
{\ddot x}^s_{(l)} + {\tilde \Gamma}_{ij}^s {\dot x}^i_{(l)}{\dot
x}^j_{(l)} + \frac{\pi}{T_{(l)}}{\tilde g}^{(l)ks} \Bigl( \frac
{\partial {\hat B}^{(l)}_{ij}}{\partial x^k_{(l)}} - \frac
{\partial {\hat B}^{(l)}_{ik}}{\partial
x^j_{(l)}}\Bigr)X^i_{(l)}{\dot x}^j_{(l)} $$
  $$
-\frac{\pi}{T_{(l)}}{\tilde g}^{(l)ks}{{\hat B}^{(l)}_{ki}}\bigl(
{\dot X}^i_{(l)} - {\dot X}^i_{(l)}\Big|_{t_{(l)}=0} \bigr) = 0;
\eqno(7)
  $$
here, $\Gamma_{ij}^s$ and ${\tilde \Gamma}^{(l)s}_{ij}$ are
symmetrical connections (see, e.g., Ref. 8)for the particle and
for the $l$th excitation, respectively; indices $i, j$ and $s$
take the values 1, 2, 3.

Let us analyze Eqs. (6) and (7). First of all, let us specify the
form of tensors $g_{ij}$ and ${\tilde g}^{(l)}_{ij}$ in
three-dimensional space. We have noted that the deformation of the
vacuum medium in the neighborhood of the particle can be related
to the availability in the particle of deformation (gravitational)
potential, which we shall designate as $V(r)$. If we take $V(r)$
to mean the potential of the particle related to its self-energy,
then a nonparametrized metric tensor in this region of the space
will take the form
  $$
g_{ij} ={\rm const} \ M \delta_{ij} (1-V(M; r)),
  $$
where the first term corresponds to the coefficient of metric of
flat space, that is, of an undistorted vacuum medium
($g_{ij}\rightarrow \delta_{ij}$). Similarly, for the $l$th
excitation
  $$
{\tilde g}_{ij}^{(l)} = {\rm const} \  m_{(l)} \delta_{ij} (1-
W_{(l)}(m_{(l)}; r)); \eqno(9)
  $$
here, $m_{(l)}$  is the mass of an $l$th excitation (the quantity
$m_{(l)}$, i.e., an additional deformation, is induced on the
$l$th superparticle when the particle strikes it). Because the
inequality $m_{(l)}\ll M$ holds, $W_{(l)}$ is local, but it is
identical by its nature to potential $V$. Differences between them
consists of the value of mass coefficient: for the particle,
$V\propto M$; for the elementary excitation, $W_{(l)} \propto m$.

Clearly, from the physical point of view, the motion of the metric
${\tilde g}_{rj}^{(l)}\rightarrow {\tilde g}^{(l)}_{r+\alpha, j}$
under the effect of operator ${\hat A}^{-1}$ should have a
definite duration in time. Then the process  of "knocking out" by
the particle from the vacuum medium of the next $l$th excitation
can  be conventionally subdivided into three stages:

\vspace{2mm} \noindent (1) The union of the particle with the
$l$th oncoming superparticle and the formation of a common system.

\vspace{2mm} \noindent (2) As a result of stage (1), the
intensification of internal processes in the $l$th superparticle
occurs, which processes continue during a definite time interval.

\vspace{2mm} \noindent(3) Then the decay of the system occurs,
that is, the further motion of the particle and the $l$th
excitation emitted by the respective superparticle.

\vspace{2mm} It is reasonable to assume that at the first stage
the interaction operator in Eqs. (6) and (7) is still not engaged,
${\hat B}^{(l)}_{ij} = 0$. Then the termwise difference between
Eqs. (6) and (7) is reduced to the form
   $$
({\ddot X}^s_{(l)} -{\ddot x}^s_{(l)}) + (\Gamma^s_{ij}{\dot
X^i_{(l)}}{\dot X^j_{(l)}} - {\tilde  \Gamma}^s_{ij}{\dot
x^i_{(l)}}{\dot x^j_{(l)}}) = 0, \eqno(10)
   $$
and Eq. (10) can be considered as the equation determining the
point of intersection of geodesics particle and excitation. The
particle and the excitation are united here to a common system,
and because of this the acceleration that one of the partners of
the system experiences coincides with the acceleration that the
other partner experiences. Therefore, the difference in the first
set of parentheses in Eq. (10) is equal to zero, and we obtain the
relation
  $$
\Gamma^s_{ij}{\dot X}^i_{(l)}{\dot X}^j_{(l)} = {\tilde
\Gamma}^{(l) s}_{ij}{\dot x}^i_{(l)}{\dot x}^j_{(l)}. \eqno (11)
  $$

   The structures of fields $\Gamma^s_{ij}$ and  ${\tilde
\Gamma}^{(l) s}_{ij}$ in the point of cross-geodesics are
identical. However, the values of intensity of the fields are
different; coefficients  $\Gamma^s_{ij}$ are generated by the mass
$M$ [according to (8)], $\Gamma \propto \partial g/\partial X
\propto V(M)\propto M]$, but coefficients ${\tilde \Gamma}^{(l)
s}_{ij}$ are generated by the mass $m_{(l)}$ [according to (9),
${\tilde \Gamma}^{(l) s} \propto W_{(l)}(m_{(l)})\propto
m_{(l)}$]. Hence the relation $\Gamma^s_{ij}/{\tilde \Gamma}^{(l)
s}_{ij}= M/m_{(l)}$ holds.

  Let us designate the velocity of the particle at the point of
emission of the $l$th excitation as $v_{0l}$, and let us assume
the rate of excitation at this point (initial rate) is equal to
the velocity of light, $|{\dot x}_{(l)}|_0=c$. Then, instead of
expression (11), we obtain
  $$
Mv^2_{0l}=m_{(l)}c^2.  \eqno(12)
  $$
As seen from relation (12), we face here an usual situation: the
particle creates at each collision a virtual excitation in the
vacuum with the kinetic energy equal to the kinetic energy of the
particle itself.

In order that such solutions take place, we must postulate the
following properties of the vacuum medium: first, the shape and
size of superparticles can fluctuate, and the superparticles are
less rigid as compared with the particle; second, by analogy with
the behavior of atoms in a solid, let us assume that
superparticles participate in collective vibrations at which they
periodically shift from the initial equilibrium states.

Having accepted the present hypotheses, we come to the mechanism
of emission of excitation. In fact, a moving particle at an
inelastic contact with "soft" $l$th superparticle of course
deforms the latter. At a farther displacement op the particle to
the ($l+1$)st cell, the superparticle in the $l$th cell remains
deformed (excited). However, deformation of the $l$th cell swiftly
relaxes to the initial state as a result of fluctuations of the
elastic energy stored in collective vibrations of the vacuum
lattice. I this case, however, local deformation from the $l$th
cell transferred by the lattice vibrations deep into the vacuum
medium in a direction perpendicular to the trajectory of the
particle (Fig. 3), that is, the migration of the deformation in
this direction is due to elastic forces acting in the vacuum
medium lattice, but not due to the transference of particle
momentum. Thus the system in question is an open system;
therefore, expression (12) could not be considered as an energy
conservation law. The migration of this deformation can be
identified with the emission of the $l$th excitation with mass
$m_{(l)}$.

Let us note that the motion of small polaron in a solid [7] (for
the motion of the proton polaron, see Ref. 9), that of polariton
excitations or a drift of ions in a polarizable medium can be
analogous to the migration of the present elementary excitations.
In the polaron potential well at the stationary state a charge
carrier is on one of the lowest level. a shift of the particle to
the next well takes place as a result of the fluctuating filling
of the first well by the energy of the vibrating lattice, that is,
it is activated by photons. But a directed drift of the particle
is provided by the electric field, by the temperature gradient, by
he medium deformation gradient, etc. In our case the drift of
excitation in a direction perpendicular to the particle trajectory
is caused by the initial condition -- by the deformation of the
$l$th superparticle in this direction (Fig. 3).
\begin{figure}
\begin{center}
\includegraphics[scale=0.45]{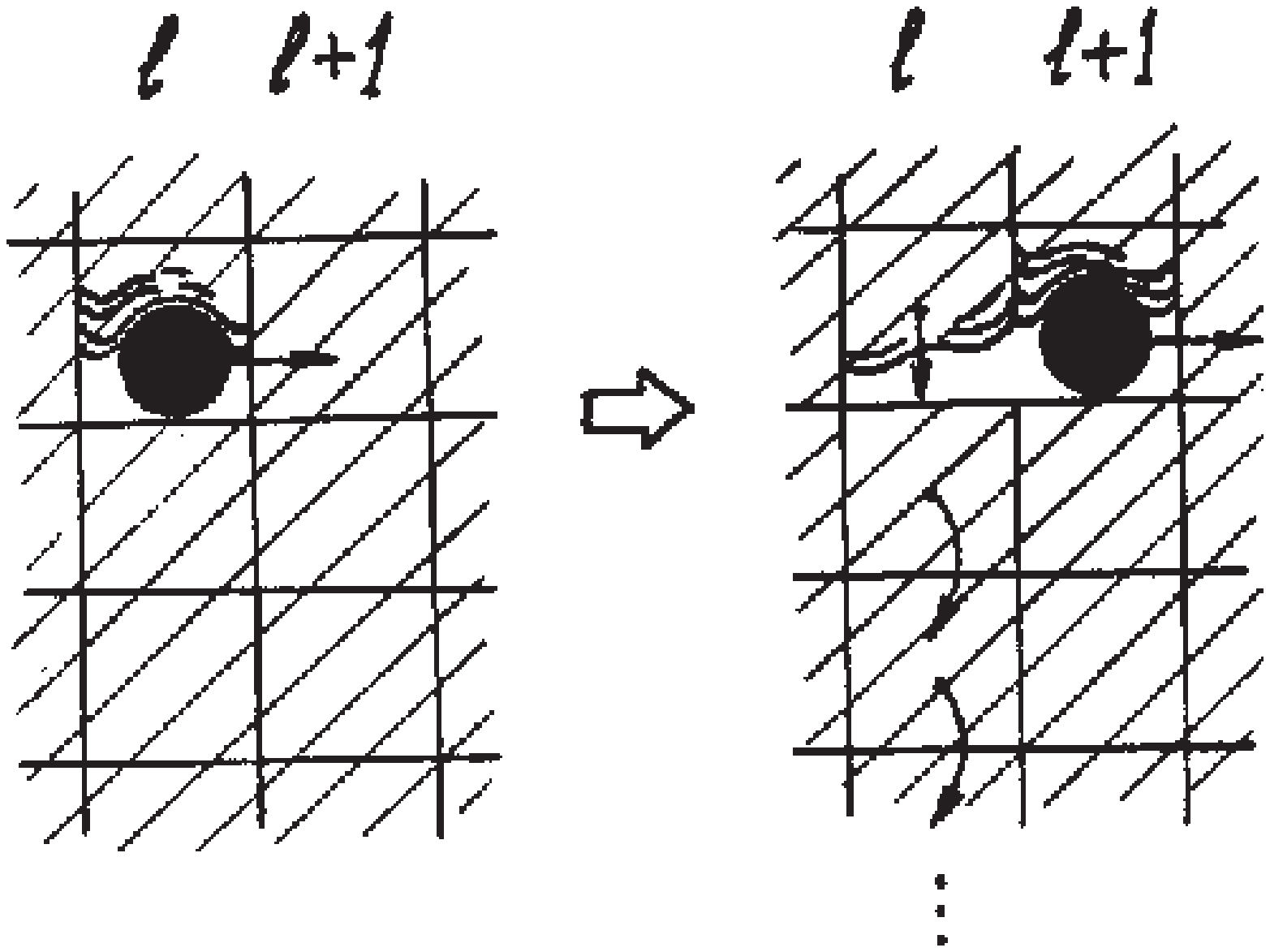}
\caption{Passage of a particle through the cell of vacuum medium
and emission of the $l$th excitation.} \label{Figure 3}
\end{center}
\end{figure}
 The realization
of the described mechanism assumes an emission of excitation at a
right angle to the trajectory of the motion of the particle.
Therefore, it is necessary to require that operator ${\hat
A}^{-1}$ (or ${\hat B}^{(l)}_{ij}$) in Eqs. (6) and (7) provides a
rotation of the space through and angle $\varphi= \pi/2$. In this
case superimposed on the axes of the original coordinate system
are the axes  of a new coordinate system in which, however, the
axes are redesignated; then the direction cosines in expression
(3) are equal to one.

The general form of Eqs. (6) and (7) is true for the particle and
the $l$th excitation only at the moment of their interaction (the
second stage), when metric tensors are influenced (also, an
external gravitational field is available). After the interaction
the particle and the $l$th excitation fly apart along their own
trajectories, where the particle at each of its path is described
by the metric tensor
  $$
g_{ij}= {\rm const} \ M \delta_{ij}, \eqno(13)
  $$
and the excitation  present in the field (8) of the particle in
each point of its path is represented by the metric tensor
  $$
{\tilde g}^{(l)}_{ij}= {\rm const} \ m_{(l)} \delta_{ij}.
\eqno(14)
  $$
Tensors (13) and (14) are constants; therefore, all their
deviations along the respective trajectories are equal to zero. As
a result, only two terms remain in Eqs. (6) and (7) -- the first
and the last ones, the latter being transformed as follows:
  $$
g^{...}{\hat B}^{(l)}_{...} \ \rightarrow \
\sqrt{m_{(l)}/M}=v_{0l}/c, \eqno(15)
  $$
  $$
{\tilde g}^{(l) ...}{\hat B}^{(l)}_{...} \ \rightarrow \
\sqrt{M/m_{(l)}}=c/v_{0l}. \eqno(16)
  $$

Here we have taken into consideration relation (12) and the fact
that by definition $g^{sk}g_{kj}=\delta^s_j$. From now on we will
omit the parentheses for indices $l$.

\vspace{4mm}
\section{Solution of the equations of motion}

\vspace{2mm} \hspace*{\parindent} Let us assume that the particle
moves along axis $X^1\equiv X$ in a Cartesian coordinate system.
The reference point of excitation radius vector $x_{l}(x^i_l)$ is
associated with the particle. Let projections of these vectors on
axes $X^1$, $X^2$, and $X^3$ be $x^1_l$, $x^2_l$, and $x^3_l$. If
we introduce a generalized coordinate $x^\perp_l =\sqrt{(x^2_l)^2
+ (x^3_l)^2}$ then the problem is reduced to a two-dimensional
one: the particle moves along axis $X$, and migrating excitations
are characterized by projections $x^\parallel_l$ and $x^\perp_l$
on the axis and on the axis perpendicular to it, respectively.
Thus the system of equations (6), (7) in Euclidean space with
regard for formulas (15), (16) takes the form
  $$
{\ddot X}_l + \frac{\pi}{T_l}\frac{v_{0l}}{c}{\dot x}_l^\perp = 0,
\eqno(17)
  $$
  $$
{\ddot x}_l^\perp - \frac{\pi}{T_l}\frac{c}{v_{0l}}({\dot X}_l -
v_{0l})= 0, \eqno(18)
  $$
  $$
{\dot x}^\parallel_l = 0.     \eqno(19)
  $$
Recall that indices by unknowns $X_l$ as well as by $x^{\perp
(\parallel)}_l$ point to their dependence on proper time $t_l$ of
the $l$th excitations. Differentiations of Eq. (18) with respect
to $t_l$ yields
  $$
{\mathop{X}\limits^{...}}^\perp_l
-\frac{\pi}{T}\frac{c}{v_{0l}}{\ddot X}_l =0.  \eqno(20)
 $$
Let us express ${\ddot X}_l$ in terms of Eq. (20), and let us
substitute it into Eq. (17). As a result, we arrive at the
equation of harmonic oscillator for $x^\perp_l$:
  $$
{\dot x}^\perp_l + (2\pi /2T_l)^2 x^\perp_l = C_{1l}.  \eqno(21)
  $$
One can see from Eq. (21) that the excitation executes harmonic
oscillations with period $2T_l$ along the axis perpendicular to
the trajectory of the particle. Let us write initial conditions
for transverse coordinate of excitations:
  $$
x^\perp\big|_{t_l=0}=0, \ \ \ \ \ {\dot x}^\perp\big|_{t_l=0}=c.
\eqno(22)
  $$
With regard to conditions (22), the solution of Eq. (21) takes the
following form:
  $$
x^\perp_l = \frac{\Lambda}{\pi}\sin(\pi t_l/ T_l);   \eqno(23)
  $$
  $$
{\dot x}^\perp_l = c \cos(\pi t_l/T_l);    \eqno(24)
  $$
here, amplitude $\Lambda_l/\pi$ corresponds to the point of
maximum separation of the $l$th excitation from the particle (Fig.
2). In this case
  $$
\Lambda_l/T_l = c.  \eqno(25)
  $$

   It should be noted that solutions $x^\perp_l$ and ${\dot
x}^\perp_l$in the form in which their are written in Eqs. (23),
(24) are true only in the interval from $t_l=0$ to $t_l= T_l$ (see
below). Let us now find $X_l$ and ${\dot X}_l$ also in the time
interval from $t_l=0$ to $t_l=T_l$. We have from Eq. (17)
  $$
{\dot X}_l = C_{2l} - \frac{\pi}{T_l} \frac{v_{0l}}{c}x^\perp_l.
\eqno(22)
  $$
Initial conditions for the particle are
  $$
{\dot X}_l (t_l + \Delta t_l)\big|_{t_l=0}={\dot X}(\Delta
t_l)=v_{0l}.  \eqno(27)
  $$
Now, we obtain from Eq. (26) with regard to Eqs. (23) and (27):
  $$
{\dot X}_l = v_{0l} [1-\sin(\pi t_l/T_l)], \eqno (28)
  $$
  $$
X_l = v_{0l}t + \frac{\lambda_l}{\pi} [\cos(\pi t_l/T_l)- 1],
\eqno(29)
  $$
where we have introduced the notation
  $$
\lambda_l = v_{0l}T_l.   \eqno(30)
  $$
It is easily seen from Eq. (28) that the velocity of the particle
is a periodic time function on interval $T_l$: \ at $t_l =0$, \
${\dot X}_l = v_{0l}$; \ at $t_l=T_l/2$, \ ${\dot X}_l = 0$; \ and
at $t_l = T_l$, \  ${\dot X}_l = v_{0l}$ again.

  Let us find the relationship between the initial velocity
$v_{0l}$ of the particle (at the instant moment of time $t_l = 0$,
where $l \neq 0$) and the initial velocity $v_0$ (at the instant
of time $t=t_l = 0$, i.e., at $l=0$). Assuming $l=0$ in Eq. (28),
we go from time $t_l$ to the proper time of the particle $t$:
  $$
{\dot X}_l(t_l + \Delta t_l) = {\dot X}(t) = v_0 [1-\sin(\pi t/
T)].    \eqno(31)
  $$
For the instant of time $t_l =\Delta t_l$, it follows from Eq.
(31) that
  $$
{\dot X}(\Delta t_l)= v_0 [1- \sin (\pi \Delta t_l/T_l)],
\eqno(32)
  $$
or, since $\Delta t_l = T l /2 N$ (see Fig. 4), we obtain instead
of Eq. (32),
  $$
{\dot X}(\Delta t_l) =v_0 [1-\sin(\pi l /2N)].     \eqno(33)
  $$
\begin{figure}
\begin{center}
\includegraphics[scale=0.6]{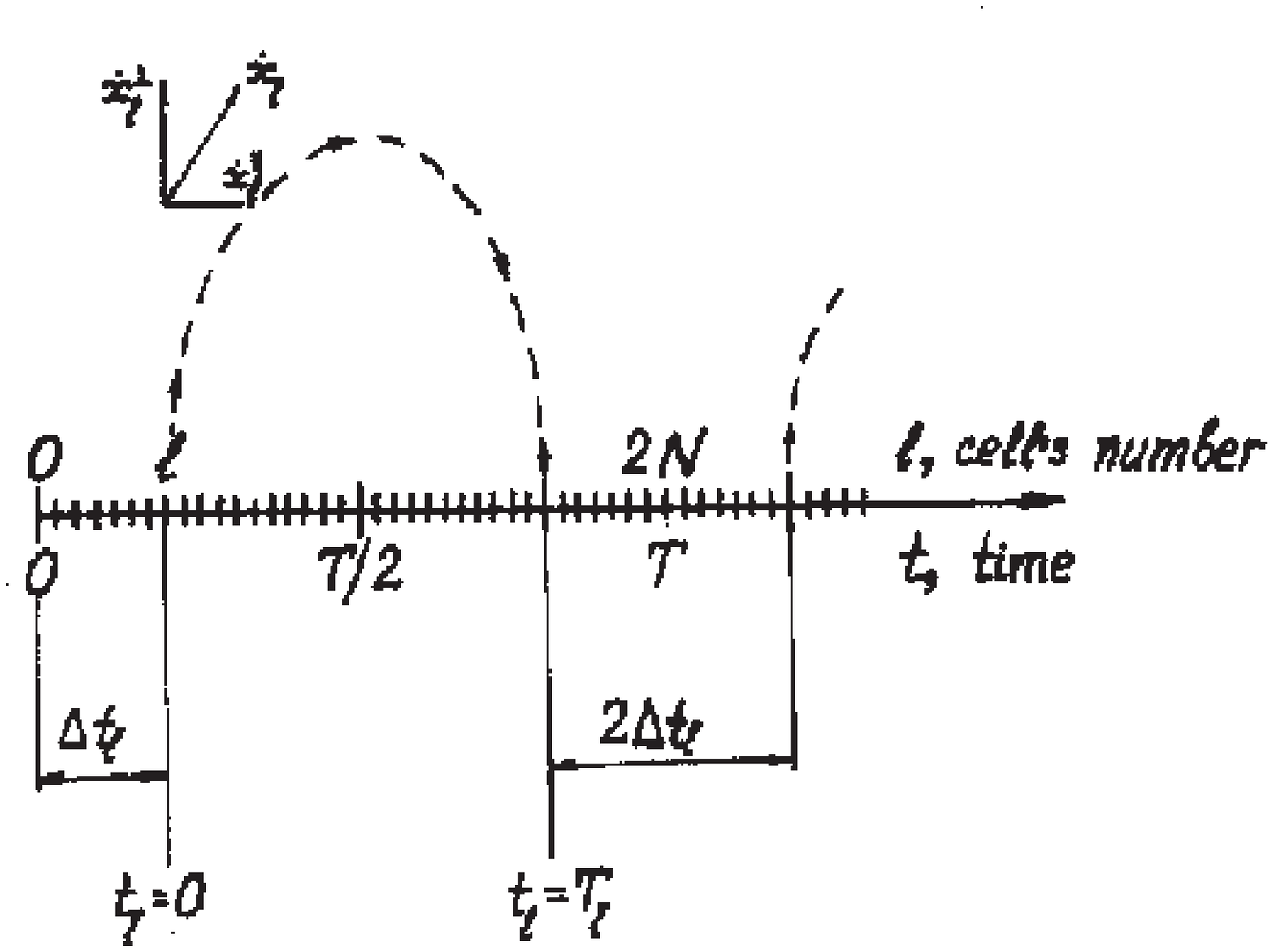}
\caption{ Trajectories of motion of the $l$th excitation. It is
emitted in time $\Delta t_l$ ($t$ and $t_l$ are proper times of
the particle and the $l$th excitation, respectively, with $t=t_l +
\Delta t_t$, where $\Delta t_l =T l / 2 N$, \ $T_l=T-2\Delta t_l
\equiv T(1-l/N)$, and $l=\overline{0,\ N-1}$).} \label{Figure 4}
\end{center}
\end{figure}
But it is at the instant of time $t=\Delta_l$ that the $l$th
excitation is emitted. Therefore, formula (33) should be compared
with formula (27). As a result, we obtain for $v_{0l}$
  $$
v_{0l}=v_0 [1-\sin (\pi l/ 2 N)].   \eqno(34)
  $$

   Thus we have obtained harmonic solutions for the velocity and
the coordinate of the particle, and in this case the motion of the
particle is characterized not only by the time half period $T_l$
of the cycle, but also by the space period $2 \lambda_l / 2\pi =
\lambda_l /\pi$. However, there is a time delay $2\Delta t_l$ (and
also a spatial delay) between the instant of absorption of the
$l$th excitation and the instant of its subsequent emission (see
Fig. 4). Therefore, to extend solutions (28), (29) as well as
(23), (24) to the following oscillations of the particle (Fig. 2),
a particle transition from the ($n-1$)st to the $n$th oscillation
should be included in these formulas, that is, a quasicyclicity in
parameter $t_l$ should be introduced. Substitution $t_l
\rightarrow t_{nl}$ meets this requirement, where
  $$
t_{nl} =t_l + 2(n-1)T_l, \ \ \ \ \ n= 1, 2, 3, ...;
  $$
  $$
T_l=T(1-l/N), \ \ \  0 \leq t_l \leq T_l, \ \ \
 l=\overline{{0,\ N-1}}.      \eqno(35)
  $$
 Thus the complete solution of system of Eqs. (17), (18) is
determined by Eqs. (23), (24) and (28), (29) in which, however,
the quasicontinuous parameter $t_{nl}$ given by formula (35) takes
the part of proper time.

    We now turn our attention to the solution of Eq. (19). It
follows from this solution that ${\dot x}^\parallel_l ={\rm
const}$. We can use the law of conservation of momentum to find
${\dot x}^\parallel_l$, since collisions of the particle with
superparticles are inelastic. However, after a time $t_l=T_l$ has
passed from the instant of emission of the $l$th excitation, the
latter should again be absorbed by the particle. The path
traversed by the particle in time $t_l =T_l$, according to Eq.
(29)is equal to
  $$
X_l(T_l)= 3\pi v_{0l} T_l/2.  \eqno(36)
  $$
Equating $X_l(T_l)$ to product ${\dot x}_l^\parallel$, we find
  $$
{\dot x}_l^\parallel = 3\pi v_{0l} /2.
  $$
To take into account that function ${\dot x}_l^\parallel$ relates
to the $n$th oscillation of the particle, it is sufficient to
introduce time parameters
  $$
\tau_{nl}=(2n-1)\Delta t_l + (n-1)T_l, \ \ \  \ \ \ n= 1, 2, 3,
...;
  $$
  $$
\Delta t_l = T_l l /2N, \ \ \ \ \ T_l =T(1-l/N), \ \ \ \ \
l=\overline{{0,\ N-1}}.  \eqno(38)
  $$
Then, instead of Eq. (37), we have for ${\dot x}_l^\parallel$  as
for a function of proper time $t$ of the particle
  $$
{\dot x}_{nl}^\parallel = \frac{3\pi}{2} v_{0l}
\Theta(t-\tau_{nl})\Theta(\tau_{nl} + T_l -t);  \eqno(39)
  $$
here, the $\Theta$ function has been introduced [$\Theta (t)=1$
when $t\geq 0$, and $\Theta (t)= 0$ when $t< 0$].

\vspace{4mm}
\section{Orthodox wave mechanics}

\vspace{2mm} \hspace*{\parindent} Let us write
 the Lagrangian (1) in two-dimensional Euclidean space in the form
  $$
L=M{\dot X}^2/2 + m\big[({\dot x}^\parallel)^2  +  ({\dot
x}^\perp)^2 \big] /2
  $$
  $$
-\frac{\pi}{T} \sqrt{mM} \ \bigl({\dot x}^\perp X + v_0 x^\perp
\bigr);   \eqno(40)
  $$
here, parameter $m$ is the mass of the whole excitations cloud
[the last three terms in expression (40) describe the motion of
the center of mass of the cloud and its interaction with the
particle];  $1/T$ is the frequency of collisions.

  Using the substitution
  $$
{\dot x}^\perp = {\dot \chi}^\perp + \pi \sqrt{M/m} \ X/T,
\eqno(41)
  $$
the Lagrangian (40) is reduced to the canonical form
  $$
L= M{\dot X}^2/2 - M \bigl(\frac{\pi}{T}\bigr)^2 X^2/2 \ \ \ \ \ \
\ \ \ \ \ \ \ \ \ \ \ \ \ \ \ \ \ \ \ \ \ \ \ \ \
  $$
  $$
+ m \bigl[({\dot \chi}^\perp)^2 + ({\dot \chi}^\parallel)^2
\bigr]/2 - \pi \sqrt{m M} \ v_0 x^\perp /T.   \eqno(42)
  $$
From this we obtain for the effective Lagrangian of the particle
  $$
L_{\rm eff} = M{\dot X}^2/2 - M \bigl(\frac{\pi}{T}\bigr)^2 X^2
/2.     \eqno(43)
  $$
As is seen, function (43) describes a harmonic oscillator. It
should be noted that this form of writing $L_{\rm eff}$ does not
take into account the translational motion of the particle;
therefore, the results based on expression (43) refer to the
behavior of the particle in the system of the particle and
excitations center of mass.

   Other terms in expression (42) characterize the effective
kinetic and potential energies of the cloud of excitations. Let us
introduce the Hamiltonian corresponding to the function $L_{\rm
eff}$:
  $$
H_{\rm eff} = p^2/2M + M(2\pi / 2 T)^2 X^2 /2. \eqno(44)
  $$

   Solutions of the equations of motion given by function (44) are
well known for different presentations (see, e.g., Refs. 10 and
11). Specifically, we can obtain from expression (44) the
Hamilton-Jacobi equation
  $$
(\partial S_1 /\partial X)^2 /2M + M (2\pi /2T)^2 X^2 /2 = E,
\eqno(45)
  $$
from which we obtain the equation for a shortened action:
  $$
S_1 = \int\limits^X  p \  d  X = \int\limits^X  \sqrt{2M \bigl[ E-
(2\pi /2T)^2 X^2 /2 \bigr]} \ d X.    \eqno(46)
  $$
It is easily find  [10,11] a solution $X$ as a function of $t$ for
the harmonic potential:
  $$
X = \frac {(2E/M)^{1/2}}{(2\pi /2T)} \sin(2\pi t /2 T). \eqno(47)
  $$

  In action-angle variables, let us determine the increment of the
action of the particle in period $2T$:
  $$
J=\oint p d X = \frac {2E}{(2\pi /2T)}\int\limits^{2\pi}_0
\cos^2\vartheta d \vartheta = E \cdot 2T = E/\nu,  \eqno(48)
  $$
here, the notation $\nu=1/2T$ and the substitution
$X=\sqrt{2E/[M(2\pi/2T)^2]} \sin\vartheta$ are used. Constant $E$
is initial energy of the particle, and so we obtain by
substituting $E=Mv_0^2/2$ into Eq. (48)
  $$
J= Mv_0 \cdot Mv_0 = p_0 \lambda.   \eqno(49)
  $$
Here, $\lambda =v_0 T$ can be called the reduced amplitude of the
oscillating particle, since, according to Eq. (47), the amplitude
of an oscillator is equal to $2Tv_0/2\pi$ [compare with expression
(30), where $\lambda$ is the reduced spatial period of
oscillations of a moving particle]. Thus the effective Hamiltonian
(44) of the particle enables formulas (48) and (49) to be
obtained, which, when we assume $J=h$ ($h$ is Planck's constant),
lead to two main relationships in quantum mechanics:
  $$
E=h\nu, \ \ \ \ \ \   p _ 0 = h / \lambda.    \eqno(50)
  $$
In formulas (50), $\nu$ is the frequency of oscillation of a
particle with initial energy $E$, and $\lambda$ is the reduced
amplitude of oscillations which we can identify with the de
Broglie wavelength.

   Let us write the complete action:
   $$
S=S_1 - E t = \int\limits^X p \ d X - E t.   \eqno(51)
   $$
Here, $S_1$ is defined in Eq. (46)and describes the particle in
the finite region where the variable $X$ is restricted by
amplitude $\lambda / \pi$. However, passing to configurational
space, we can write the coordinate of the particle, taking into
account its $n$th oscillation, in the form
  $$
X \rightarrow X_n + (n-1)\lambda /\pi;
  $$
at $n \gg l$, it turns into a continual variable. Now, when we
assume in Eq. (5), $p={\rm const}= p_0 =Mv_0$, we obtain an action
that is transformed to a characteristic function of uniform
infinite motion in configurational space:
  $$
{\widetilde S}_{\rm part} =Mv_0X - E t.  \eqno(52)
  $$
But if we assume in Eq. (51) according to expression (50) that
$p={\rm const} =h/\lambda$ and $E=h\nu$, then we obtain the action
in the form
  $$
{\widetilde S}_{\rm wave} = h(X/\lambda - \nu t),   \eqno(53)
  $$
and then ${\widetilde S}_{\rm wave}/h$ can be related to the phase
of a monochromatic wave propagating along the $X$-axis of
configurational space. Both ${\widetilde S}_{\rm part}$ and
${\widetilde S}_{\rm wave}$ are obtained from one and the same Eq.
(51) in one and the same approximation but in different
representations: the former in terms of the initial momentum $p_0$
and energy $E=p_0^2/2M$ of the particle and the latter in terms of
amplitude $\lambda$ and frequency $\nu$ of its oscillations. Both
expressions (52) and (53) are identical, and this allows us to
compare the particle moving in configurational space with the wave
of the form
  $$
\psi = a \exp \bigl( {2\pi {\widetilde S}_{\rm wave}/h}\bigr).
\eqno(54)
  $$
With regard to (30) and (50), a substitution of the wave function
(54) into the wave equation
  $$
\Delta \psi - \frac{1}{(v_0/2)^2}\frac{\partial^2 \psi}{\partial
t^2} =0
  $$
results in Schr\"odinger's equation (compare with Ref. 12, Chap.
1).

    The phase of the wave (54) carriers reliable information only
about the extremum values of the parameters. Therefore, it is
clear that the formalism of the $\psi$ function can successfully
describe only stationary states. But for dynamic variables the
$\psi$ function gives only a probabilistic estimate which is also
clear, since this formalism does not assume a mechanism describing
the system dynamics. At the same time, if the vacuum is really a
medium, then the need for a supplement or modification of action
${\widetilde S}_{\rm wave}$ is obvious. In our interpretation of
the hidden dynamics of the particle, this is the transition
${\widetilde S}_{\rm wave} \rightarrow S$ and then to the
equations of motion (17) to (19) in real space. In the considered
dynamics, the motion of a particle is determined by solutions (29)
and (30) with regard to formula (35). It comprises the time
modulation of coordinate $X(t)$ with frequency $1/T=v_0/\lambda$,
and this is why such motion can be associated with the traveling
wave [some similarity with the traveling wave is especially
obvious in coordinates ${\widetilde X}, \ {\widetilde t}$, Fig.
5].
\begin{figure}
\begin{center}
\includegraphics[scale=0.6]{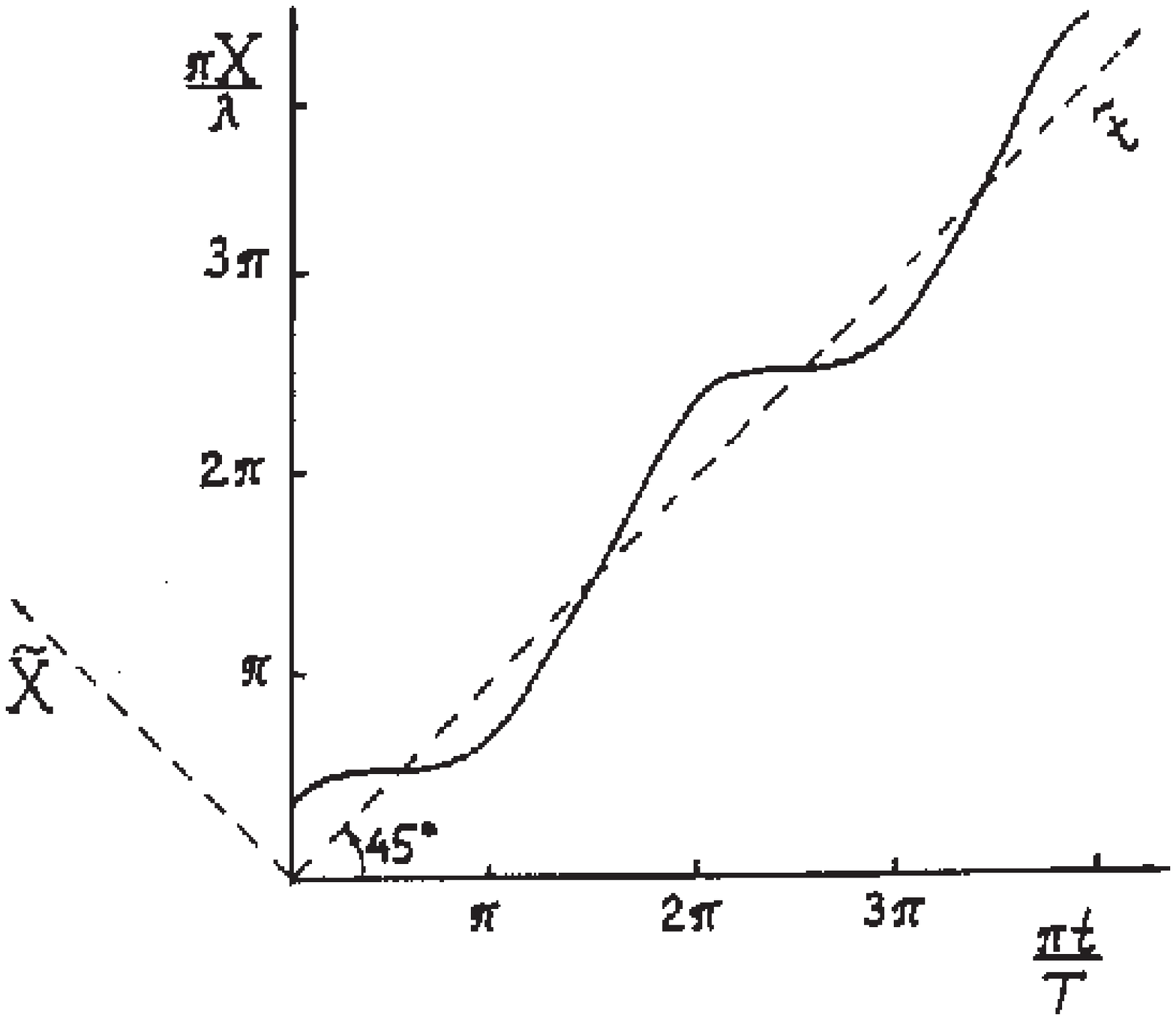}
\caption{Graph in dimensionless coordinates $\pi X_n /\lambda$ as
a function of its proper dimensionless time $\pi t_n / T$, plotted
according to Eq. (29) with regard to formula (35), that is, the
graph of solution of equation $\pi X_n/\lambda =\pi t_n /T +
\cos(\pi t_n /T)$.} \label{Figure 5}
\end{center}
\end{figure}

\vspace{4 mm}
\section{Conclusion}

\vspace{2mm} \hspace*{\parindent} In the present paper we
postulated the availability of a discrete vacuum medium and
provided it with the properties that would enable the particle to
move between superdensely packed "molecules" of a vacuum.

   In addition, the frequency of collisions of the particle with
excitations of this discrete medium can be written in two ways. On
the one hand, the frequency of each $l$th collision for a particle
$1/T_l=v_{0l}/\lambda$, according to expression (30). On the other
hand, according to expression (25), the frequency of collisions
with a particle for each $l$th excitation is $1/T_l =c/\Lambda_l$.
At $l=0$, parameters of a particle and a cloud of excitation
attain extremum values, so, by comparing expressions (30) and (25)
at $l=0$, we obtain
  $$
v_0/\lambda = c/\Lambda.  \eqno(56)
  $$
Relation (56) connects the spatial period of oscillations of a
particle $\lambda$ (i.e., the de Broglie wavelength) to amplitude
of the cloud $\Lambda$. For example, in the case of a free
electron with velocity $v_0 = 10^5$ cm/s and a de Broglie
wavelength $\lambda = h/Mv_0 6\times 10^{-5}$ cm, we obtain for
the characteristic dimension of excitation cloud according to
expression (56), $\Lambda=\lambda c/v_0 \cong 2$ cm, that is,
$\Lambda$ is a macroscopic value.

    Let us also evaluate the number of excitations $N$ in the
cloud of this electron. For this purpose, let us divide the de
Broglie wavelength $\lambda$ by the dimension of superparticle
$R_0$. If we assume $R_0\sim 10^{-28}$ cm, then we obtain $N \sim
\lambda/R_0 \sim 10^{22}$. In our model, the cloud of excitations
surrounding a particle "feels" obstacles at a distance $\sim
\Lambda$ from the particle, and those excitations transmit the
respective information to the particle. Such motion is close in
principle to de Broglie's "motion by guidance," which he
associated with a constant intervention of subquantum medium
[6,12].

    Spatial oscillations of a particle are apparently the result
of its adiabatic motion at which the particle does not leave
behind faults in the vacuum medium structure. At the same time,
the oscillator is characterized by the adiabatic invariant
$J=E\nu$. In mechanics the mass $M$ of the particle and by the
system elasticity constant $\gamma$: \ $\nu =
\sqrt{\gamma/M}/2\pi$. Parameters $M$ and $E$ are the
characteristics of the particle. Hence, assuming $J=h$, we
automatically determine Planck's constant as an adiabatic
invariant of oscillator whose elasticity constant is defined by
elastic properties of the vacuum medium. Let us note that such an
interpretation of the constant $h$ is not contradictory to that by
Lochak [13], whose studies are devoted to the physical nature of
$h$.

   In this paper we correlated the deformation of the vacuum
medium with gravitational potential. Therefore, excitations of the
vacuum medium could be called gravitons, but they basically differ
from gravitons of the general theory of relativity (where they
have polarization 2). In view of this, we have called the present
excitations, "inertons." This name seems quite acceptable, since
it reflects a connection with the motion of the particle, that is,
with inertia.

   The experimental detection of inertons would be a direct
confirmation of the existence of a physical vacuum in the form of
a peculiar quantum medium. This, in turn, would be a positive
argument in favor of the concept of the oscillating motion of
elementary particles set forth in this paper.

\vspace{6mm} \noindent
 {\bf Acknowledgment}

\vspace{2mm}
 The authors are very grateful to Mr. Mieczys{\l}aw Grudzie\'n for
financial help, which afforded us the opportunity to publish this
paper.


\newpage
\begin{thebibliography}{99}

\bibitem{1} A.A. Grib, {\it Problem of Vacuum Noninvariance in
           Quantum Field Theory} (Atomizdat, Moscow, 1978) (in
           Russian).

\bibitem{2} L.B. Okun, {\it Physics of Elementary Particles}
           (Nauka, Moscow, 1988) (in Russian).

\bibitem{3} V.A. Dubrovsky, Dok. Akad. Sci. USSR {\bf 282}, 83
            (1985) (in Russian).

\bibitem{4} S.W. Hawking, in {\it General Relativity}, edited by
           S.W. Haking and W. Israel (Cambrodge University Press,
           1979), p. 746.

\bibitem{5} P.I. Fominm in {\it Quantum Gravity}, Proceedings of
            the Fourth Seminar on Quantum Gravity, Moscow, USSR,
            1987, edited by V. Markov, V. Berezin, and V.P. Frolov
            (World Scientific, Singapore, 1988), p. 813.

\bibitem{6} L. de Broglie, Ann. Fond. L. de Broglie {\bf 12}, 399
            (1987).

\bibitem{7} Yu.A. Firsov, editor, {\it Polarons} (Nauka, Moskow,
            1975) (in Russian).

\bibitem{8} B.A. Dubrovin, S.P. Novikov, and A.T. Fomenko,
            {\it Modern Geometry: Methods and Applications}
            (Nauka, Moscow, 1986) (in Russian).

\bibitem{9} V.V. Krasnoholovets, M.A. Protsenko, and P.M. Tomchuk,
            Int. J. Quant. Chem. {\bf 33}, 349 (1988).

\bibitem{10} H. Goldstein, {\it Classical Mechanics} (Nauka, Moscow, 1975)
             (Russian translation).

\bibitem{11} D. ter Haar, {\it Elements of Hamiltonian Mechanics}
             (Nauka, Moscow, 1974) (Russian translation).

\bibitem{12} L. de Broglie, {\it Heisenberg's Uncertainty Relations and
             the Probabilistic Interpretation of Wave Mechanics}
             (Mir, Moscow, 1986) (Russian translation).

\bibitem{13} G. Lochak, in {\it Didaktik der Physik Vortage}, Her.
             W. Kuhn (Deutsche Physikalische Gesellschaft, Physikertagung,
             Berlin, 1987), S. 607.

\end{thebibliography}
\end{document}